\documentclass[longbibliogrphy,sectionnumbering]{beilstein}

\nolinenumbers
\begin{document}

\title{
Circular dichroism of chiral Majoranas
} 

\author{Javier Osca}
\affiliation{Institut de F\'{\i}sica Interdisciplin\`aria i de Sistemes Complexos
IFISC (CSIC-UIB), E-07122 Palma de Mallorca, Spain}
\author*{Lloren\c{c} Serra}{llorens.serra@uib.es}
\affiliation{Institut de F\'{\i}sica Interdisciplin\`aria i de Sistemes Complexos
IFISC (CSIC-UIB), E-07122 Palma de Mallorca, Spain}
\affiliation{Departament de F\'{\i}sica,
Universitat de les Illes Balears, E-07122 Palma de Mallorca, Spain}
\maketitle

\date{November 2, 2017}

\begin{abstract}
\background
Majorana states in condensed matter devices may be of a localized nature, such as in 
hybrid semiconductor/superconductor nanowires, or chirally propagating along the edges  such as in  
hybrid 2D quantum-anomalous-Hall/superconductor structures.
\results
We calculate the circular dichroism due to chiral Majorana states in  a hybrid structure made of a  quantum-anomalous-Hall-insulator and a superconductor.
The optical absorption of chiral Majoranas is characterized by equally spaced absorption peaks
of both positive and negative dichroism. In the limit of a very long structure
(a 2D ribbon) peaks of a  single sign are favored.
\conclusion circular-dichroism spectroscopy of chiral Majoranas is suggested as a relevant probe for these peculiar states of topological matter.
\end{abstract}



\section{Introduction}

The physics of Majorana states in condensed matter devices is attracting strong interest since a few years ago
\cite{Nayak,Alicea,StanescuREV,Beenakker,Franz,Elliott,Aguado,Lutrev}. 
The measured  zero-bias conductance peaks in hybrid semiconductor/superconductor nanowires have been attributed to the presence of localized Majorana modes on the two ends of the nanowire
\cite{Lutchyn,Oreg,Mourik,HaoZ,Deng,Das}. 
These peculiar pairs of states may be seen as nonlocal split fermions, protected by an energy gap that separates them from other  normal states lying at finite energies. Besides the zero energy of the Majorana state, also the peak height was recently seen to coincide with the expected value $2e^2/h$ \cite{HaoZ2}.

Majorana end states in (quasi) 1D nanowires are inherently localized. By contrast, propagating Majorana
states can be formed at the edges of 2D-like hybrid structures. We refer, specifically, to the 
hybrid devices of Ref.\ \cite{He294}, consisting of a quantum anomalous Hall
insulator and a superconductor material. In such systems, chiral Majorana modes propagating along the 
edges in a clockwise or anticlockwise manner, depending on the orientation of a perpendicular magnetic field, are formed 
at the 2D interfaces between the quantum anomalous Hall
and the superconductor materials
\cite{Qi10,Chung,Wang15,Lian16,Kalad}. 
Each chiral Majorana contributes $0.5\,e^2/h$  to the linear conductance of the device, 
such that by tuning the number of Majoranas the conductance takes values $0.5\,e^2/h$ and $1\,e^2/h$ 
for the topological phases with one and two chiral Majoranas, respectively. 
 
In this work we discuss the connection between chiral Majoranas and optical absorption. We expect that in presence of chiral Majoranas, the optical absorption of circularly polarized light will differ for clockwise and anti-clockwise polarizations. The difference, known as the {\em circular dichroism} (CD) \cite{Eisfeld,Longhi}, can thus be seen as a measure of the existence of such chiral states. We want to investigate how this behavior is actually realized by explicit calculations of the optical aborption. In  previous works we analyzed the optical absorption of
localized Majoranas in nanowires \cite{Ruiz,Osca2015}. 
In those systems the CD vanishes and the presence of the Majorana is signaled by a lower-absorption plateau, starting at mid-gap energy, 
of the $y$-polarized signal with respect ot the $x$-polarized one.
It is also worth mentioning that alternative techniques for detecting Majorana fermions, based on 
microwave photoassisted tunneling in Majorana nanocircuits have been suggested in Ref.\ \cite{Dart}.

For chiral Majoranas in a 2D square or rectangular geometry the CD 
at low energies
is characterized by a sequence of equally spaced peaks, corresponding to transitions from
negative to positive energy Bogoliubov-deGennes quasiparticles.
In the usual energy ordering of quasiparticle states ($n=\pm1, \pm2, \dots$),
the selection rules are: a) transitions between conjugate states $-n\to n$
are forbidden by electron-hole symmetry, b) transitions $-n\to m$ 
are allowed only when $n$ and $m$ are both even or both odd.
The rationale behind rule b) is the constructive interference
of the corresponding quasiparticle states connected by the excitation operator
on the edges of the system.
Furthermore, it will be shown below that the
CD peaks corresponding to those even-even or odd-odd  quasiparticle transitions may be either
positive or negative. In the limit of a long 2D ribbon there is a preferred CD sign, depending on
the magnetic field orientation.
For a disc geometry the generalized angular momentum $J_z$ becomes a good quantum number.
Then, the combination of circular and particle-hole symmetries in a disc causes a vanishing absorption 
for $p_x\pm ip_y$ fields and, obviously, also a vanishing CD.  

\section{Model}

We use the model of Ref.\ \cite{He294}
for a quantum-anomalous Hall (3D) thin film in contact with two different superconductors.
This model represents 
the device as two surfaces with a certain interaction between them, with Majoranas being located at their edges.
In a Nambu spinorial representation that groups 
the field operators in the top $(t)$ and bottom $(b)$ layers,
$\left[
(
\Psi^t_{k\uparrow},
\Psi^t_{k\downarrow},
\Psi^{t\dagger}_{-k\downarrow},
-\Psi^{t\dagger}_{-k\uparrow}
),
(
\Psi^b_{k\uparrow},
\Psi^b_{k\downarrow},
\Psi^{b\dagger}_{-k\downarrow},
-\Psi^{b\dagger}_{-k\uparrow}
)\right]^T
$,
the Hamiltonian reads
\begin{eqnarray}
{\cal H} &=& 
\left[\, m_0 + m_1 \left(p_x^2 +p_y^2\right)\, \right] \tau_z\, \lambda_x 
+ \Delta_B\, \sigma_z - \mu\, \tau_z\nonumber\\
&-& \alpha\, \left(\,p_x\sigma_y-p_y\sigma_x\right)\, \tau_z\,\lambda_z \nonumber\\
&+& \Delta_p\, \tau_x + 
\Delta_m\, \tau_x\,\lambda_z\, .
\end{eqnarray}
This Hamiltonian is acting in the combined  position-spin-isospin-pseudospin space. Spatial positions are 
a treated as a 2D continuum ($xy$) and a discrete two-valued pseudospin ($z$). The two-valued spin, isospin and pseudospin degrees of freedom are represented by $\sigma$, $\tau$ and $\lambda$ Pauli matrices, respectively. The pseudospin ($\lambda$) is modeling a coupled bilayer system in which quasiparticles move.  
The set of Hamiltonian parameters is $m_0$, $m_1$, $\Delta_B$, $\mu$, $\alpha$, $\Delta_p$ 
and $\Delta_m$. The latter two are given in terms of the pairing interaction in the two layers, $\Delta_1$ and $\Delta_2$, by
\begin{equation}
\Delta_{p,m} = \frac{\Delta_1\pm\Delta_2}{2}\; .
\end{equation}

Below we numerically determine the eigenvalues and eigenstates of ${\cal H}$
using a 2D grid for $x$ and $y$. Increasing $\Delta_B$ the spectrum of low-energy 
eigenvalues evolves from a gapped (void) spectrum around zero energy at low $\Delta_B$'s, to the 
emergence of chiral near-zero-energy modes for sufficiently large values of $\Delta_B$. When the pairing parameters for each layer are equal ($\Delta_m=0$)  chiral Majoranas appear in pairs ($0-2-\dots$), while for sufficiently 
different parameters it is $\Delta_m\ne 0$ and there may be phases with odd numbers of chiral Majoranas as well.

The numerical results shown below are given in an effective unit system, characterized by the choice of 
$\hbar\equiv 1$, mass $m\equiv 1/2m_1 \equiv 1$ and a chosen length unit
$L_U$, typically 
$L_U\approx 1\, \mu{\rm m}$. 
The corresponding energy unit is then $E_U=\hbar^2/mL_U^2$.

\begin{figure}
\includegraphics[width=8.75cm,trim=1.5cm 15.cm 3.cm 2.5cm,clip]{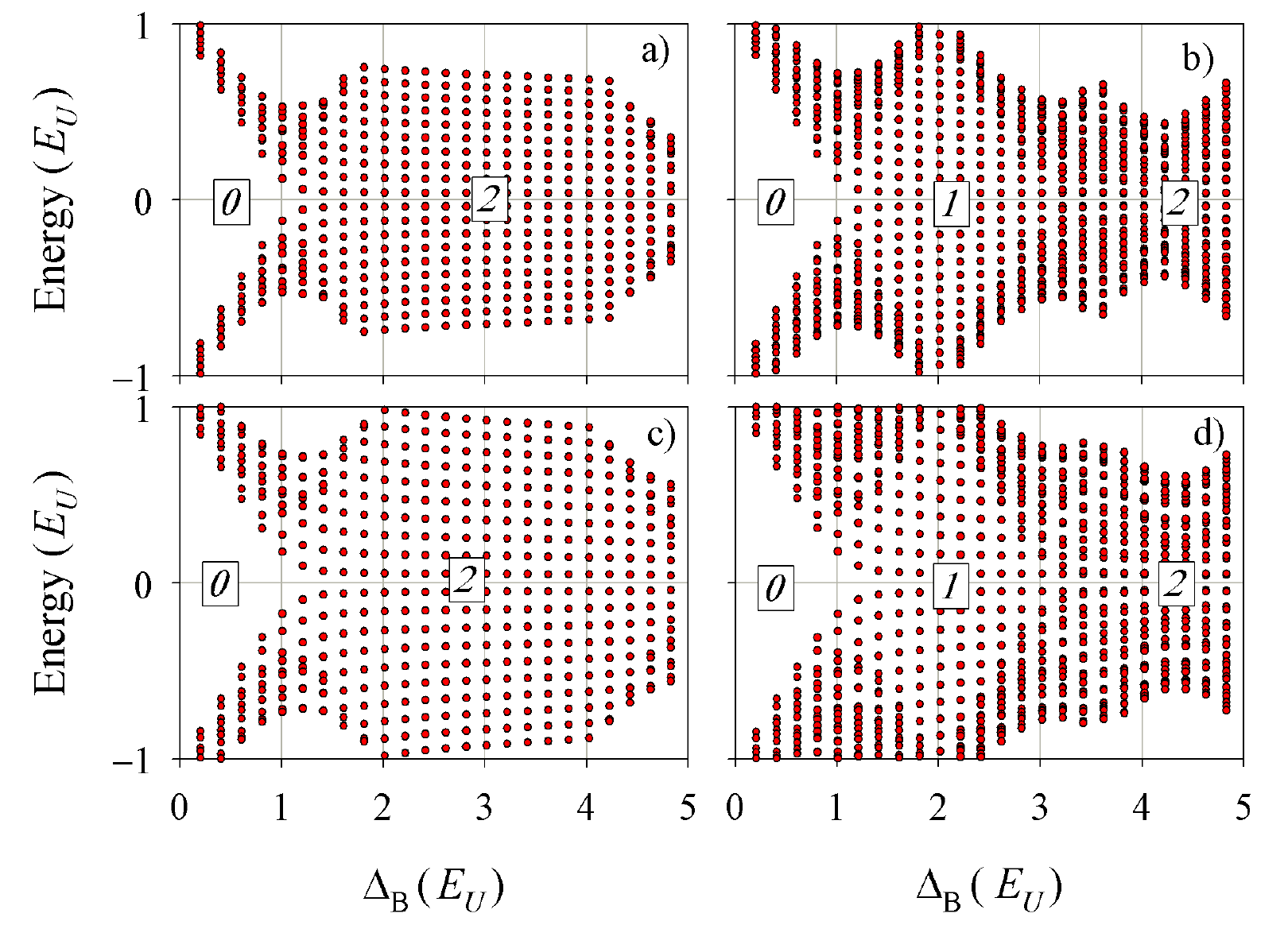}%
\caption{Energy eigenvalues closer to zero energy as a function of $\Delta_B$.
Panels a) and b) 
are for a square of dimensions $L_x=L_y=10\, L_U$, while c) and d)
correspond to a rectangle of $L_x=2 L_y=20\, L_U$. In a) and c) 
the same pairing energy is assumed in each layer 
$\Delta_1=\Delta_2=E_U$ while in b) and d) it is $\Delta_1=\Delta_2/3=E_U$.
The framed labels indicate the degeneracy of the near-zero energy states, 
which indicates the topological phase. Other parameters:
$m_0=0$, $\mu=0$, $\alpha=E_U L_U$.
}
\label{F1}
\end{figure}

\subsection{Circular dichroism}

We compute the optical absorption cross section for right ($+$) and left ($-$) circularly-polarized light from 
\begin{equation}
{\cal S}_\pm(\omega)= 4m_1^2 \sum_{k>0, s<0}{ 
\frac{1}{\omega_{ks}}\, 
\left|
\rule{0cm}{0.35cm}
\,\langle k\vert p_x\pm ip_y \vert s\rangle\,\right|^2 
\, 
\delta(\omega-\omega_{ks}) }\; ,
\end{equation}
where $\hbar\omega_{ks}=\varepsilon_k-\varepsilon_s$ is the energy difference between 
particle (unoccupied) and hole (occupied) states. The prefactor $4m_1^2$ gives the squared 
inverse effective mass ($1/m_{\rm eff}^2$) of the Hamiltonian and fixes the dimensions of ${\cal S}$
as an area.
The circular dichroism at a given frequency ${\cal S}_{CD}(\omega)$ is then defined as
the difference between the absorptions for the two circular polarizations,
\begin{equation}
{\cal S}_{CD}(\omega)={\cal S}_+(\omega)-{\cal S}_-(\omega)\; .
\end{equation}
Obviously, in absence of any chirality preference ${\cal S}_{CD}$ exactly vanishes.

\begin{figure}
\includegraphics[width=8.75cm,trim=2cm 15.cm 2.5cm 2.cm,clip]{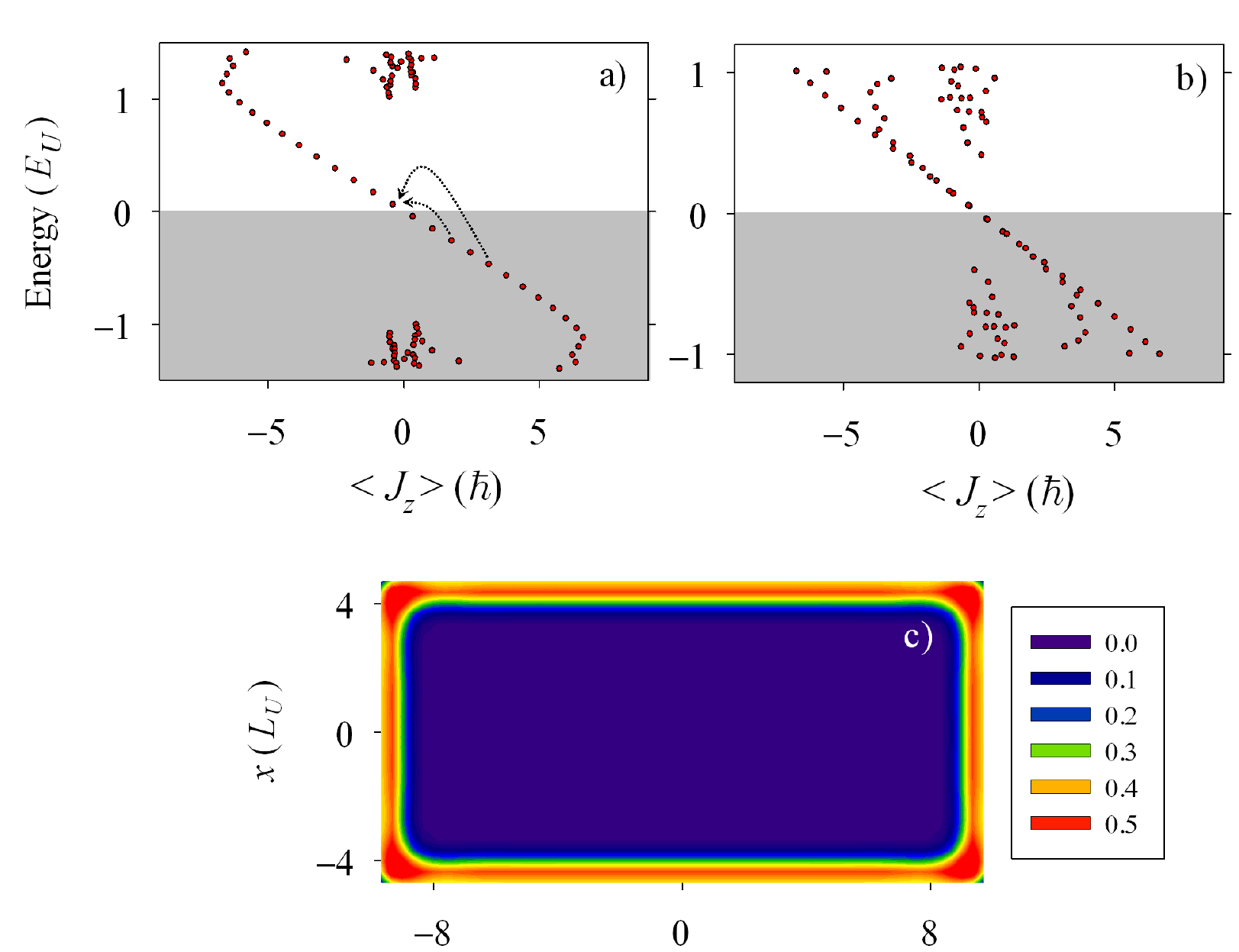}%
\caption{Energy eigenvalues as a function of $\langle J_z\rangle$. Panels a) 
and b) correspond to the phases in Fig.\ \ref{F1}d with one ($\Delta_B=2E_U$) and two ($\Delta_B=4.75 E_U$) Majorana states, respectively. 
The grey-shaded zones indicate the occupied (hole) states while 
the arrows in panel a) show the two lowest allowed transitions to the first particle state.
Panel c) shows the probability density corresponding to the lowest positive-energy state 
in panel a), adding all spin, isospin and pseudospin contributions.}
\label{F2}
\end{figure}

\section{Results and discussion}

\subsection{Chiral bands}

Figure \ref{F1} shows the evolution of the eigenvalue 
spectrum as a function of the magnetic field parameter $\Delta_B$.
The results reproduce already known results \cite{He294}. At vanishing $\Delta_B$ the spectrum around zero energy is gapped, a gap that tends to close
when increasing $\Delta_B$
by the appearance of a quasi-continuum distribution
of eigenvalues. These low-energy states are indicating the presence of propagating Majoranas, energy discretized due to the finite size of the system.
When $\Delta_1=\Delta_2$ (panels a and c) the degeneracy is such that the Majorana branches 
appear in pairs. We also notice that there is no
qualitative difference in the eigenvalue distribution between a square
and a rectangle (upper vs lower panels). It is remarkable that when a Majorana phase is well developed
the set of low energy states are equally spaced in energy. This is particularly
clear for $2<\Delta_B/E_U<4$ in panels a and c, corresponding to the phases with 
2 Majoranas. It can also be seen in panels b and d for the phases with one Majorana
while, in these panels, the equally spaced distribution is also hinted for the beginning of the phase with 2 Majoranas. 

The chiral character of the gap-closing Majorana states is clearly seen in Fig.\ \ref{F2}. The equally spaced states at low energy arrange themselves on a line
(a chiral band) when plot as a function of the $z$-component of the angular momentum. For positive $\Delta_B$ the angular momentum decreases with increasing energy, 
causing empty (particle) states to have negative $\langle J_z\rangle$, while occupied (hole) states
have positive $\langle J_z\rangle$. The results of Fig.\ \ref{F2}a,b correspond to the rectangle with different
pairing energies in each layer shown in Fig.\ \ref{F1}d. For $\Delta_B=2 E_U$ (\ref{F2}a) there is a single 
chiral band, while for $\Delta_B=4.75 E_U$ (\ref{F2}b) there are two overlapping bands. Notice that the 
overlap of states in Fig.\ \ref{F2}b degrades as the energy deviates from zero, indicating that the second Majorana band is not yet fully settled for this particular $\Delta_B$. Additionally, Fig. \ref{F2}c explicitly shows the edge character of the states of a chiral Majorana band. A similar distribution is obtained for all the 
states in a chiral band.

\begin{figure}
\includegraphics[width=8.5cm,trim=1cm 5.cm 3.5cm 4.5cm,clip]{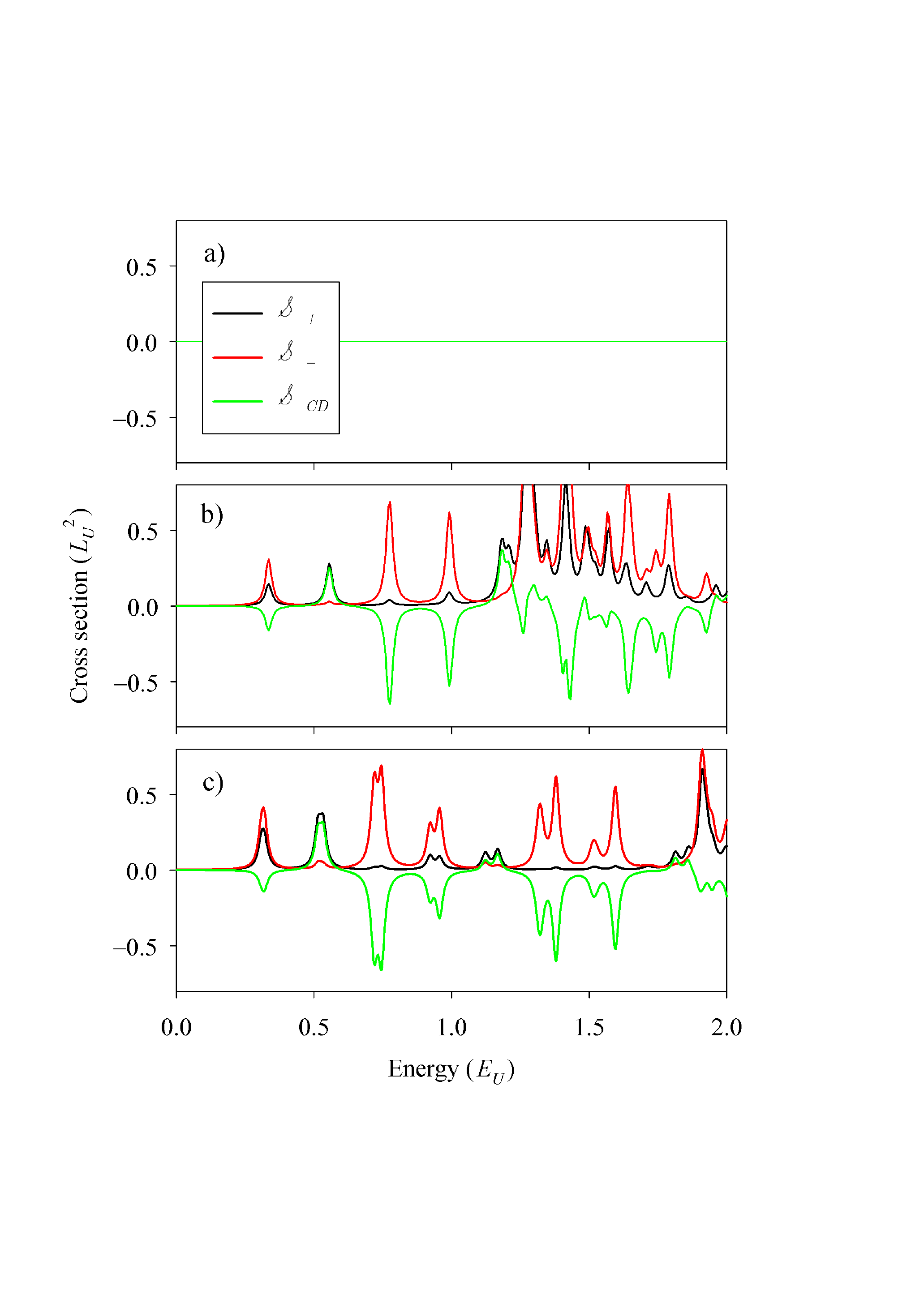}%
\caption{Absorption cross sections ${\cal S}_+$, ${\cal S}_-$ and ${\cal S}_{CD}$ defined in the 
main text. The shown results correspond to the spectra of Fig.\ \ref{F1}d for Zeeman parameters
of $\Delta B=0.3 E_U$ (a),  $2 E_U$ (b), and $4.75 E_U$ (c).}
\label{F3}
\end{figure}

\subsection{Absorption and CD}

The absorption cross sections and the CD for the spectra of the rectangle with different pairing energies in the two layers (Fig.\ \ref{F1}d) are shown in Fig.\ \ref{F3} for selected values of $\Delta_B$. They  correspond to zero (\ref{F3}a), one (\ref{F3}b) and two (\ref{F3}c) chiral bands. As anticipated, in presence of the chiral states the system develops a clear CD. The negative CD peaks dominate, due to the negative slope of the chiral bands (Fig.\ \ref{F2}a,b). It is remarkable, however, that a few positive peaks are also present. We attribute them to the fact that in a rectangular geometry $J_z$ is not a good
quantum number and, therefore, there are states with mixed angular momentum. We have performed calculations in a circular geometry confirming this interpretation. Therefore, quasiparticle scattering by the corners
plays a nontrivial role on the absorption by chiral edge states.

\begin{figure}
\includegraphics[width=8.5cm,trim=1cm 11.cm 3.5cm 4.5cm,clip]{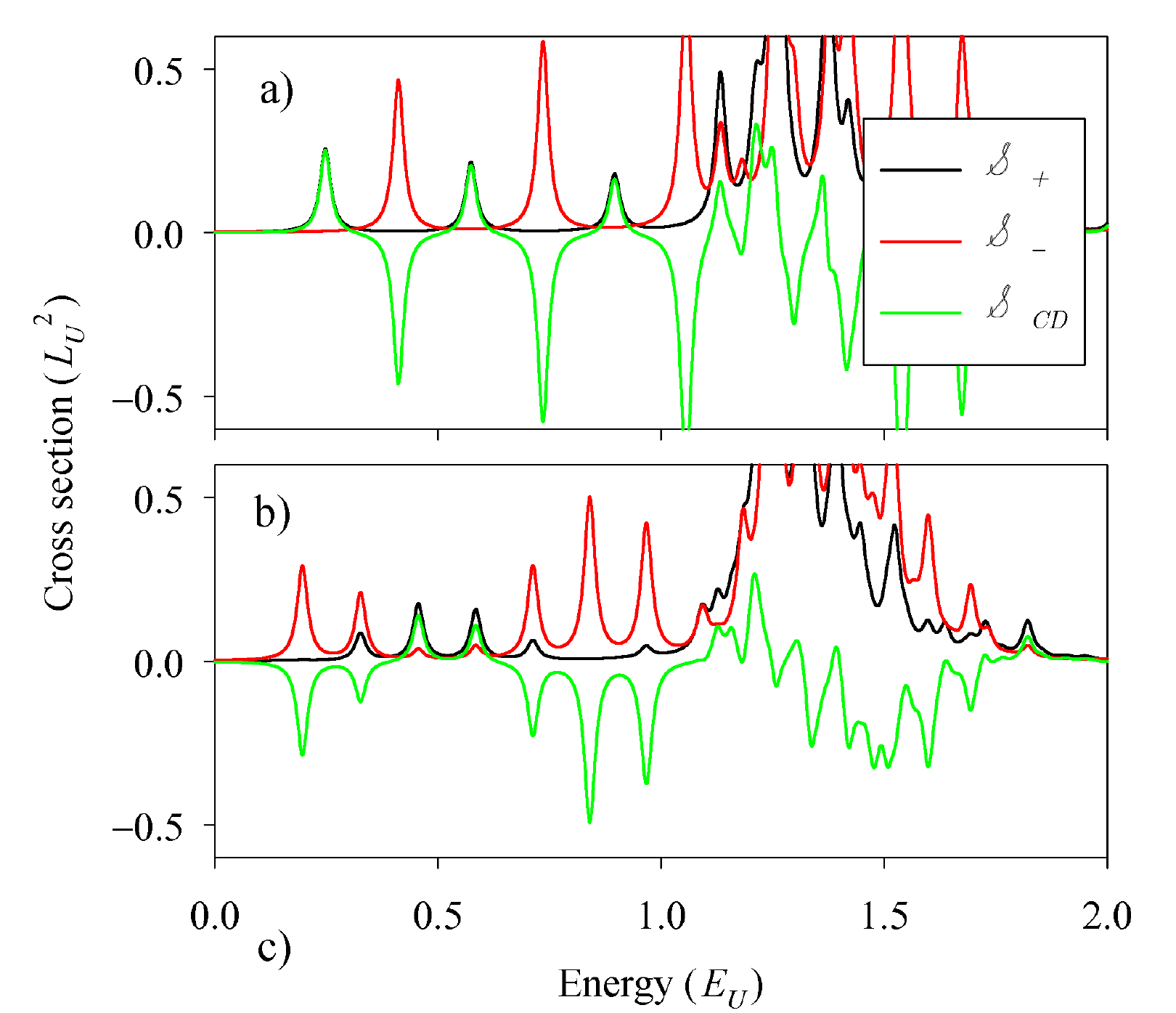}%
\caption{Absorption cross sections ${\cal S}_+$, ${\cal S}_-$ and ${\cal S}_{CD}$ 
 for a square of $L_x=L_y=20\, L_U$ (a), and for a rectangle of $L_x=6\, L_y= 60\, L_U$. In both cases 
 we used $\Delta_B=2\, E_U$ and $\Delta_1=\Delta_2/3= E_U$.}
\label{F4}
\end{figure}

The most conspicuous feature of Fig.\ \ref{F3}b is the regular energy spacing of the first few CD peaks.
Analysing them in terms of energy transitions of the chiral band it is easily  noticed
that they correspond to jumps of 3, 5, 7, \dots steps (see arrows in Fig.\ \ref{F2}a). We explain this selection rule noticing the  following restrictions for transitions from the negative $n$-th state to the positive $m$-th state 
($-n\to m$):
\begin{itemize}
\item[a)] transitions between conjugate states $-n\to n$ are forbidden by particle-hole symmetry \cite{Ruiz}, 
\item[b)] $n$ even to $m$ odd transitions (or vice versa) are forbidden because of destructive interference along the nanostructure perimeter with the excitation operator.
\end{itemize}

For a disc, $J_z$ becomes a good symmetry and, by angular momentum conservation with a dipole 
operator only the transition $-1\to1$ is possible. However, this transition is blocked by rule a)
and, therefore, no dipole absorption is possible and the CD exactly vanishes. We have also checked this behavior by explicit calculation for a device with circular geometry. For a square and rectangle, quasiparticle scattering by the corners plays a nontrivial role yielding the mentioned deviations with respect to the disc.

The pattern of equally spaced peaks is fulfilled only when one or several chiral bands are fully developed and they exactly overlap. In Fig.\ \ref{F3}c we see that the slight degradation of the two-band overlaps
of Fig.\ \ref{F2}b manifests in a small twofold splitting of the CD peaks. It is also worth stressing that once the chiral bands are fully formed, the energy positions of the first few CD peaks become $\Delta_B$-independent (cf.\ Figs.\ \ref{F2}b and \ref{F2}c).

Figure \ref{F4} shows the absorption results for different geometries, a square (\ref{F4}a) and a 
long rectangle resembling a 2D ribbon (\ref{F4}b). For the square, the first CD peaks alternate sign in a
remarkable way. On the other hand, for the ribbon the alternation is of a longer period, the positive 
peaks having a much lower intensity than the negative ones and there are groups of a few consecutive negative peaks. The 2D ribbon shape thus favors the observation of CD peaks of the same sign.  

\section{Conclusions}

In this work we have investigated the manifestation of chiral Majorana modes in the CD of the dipole absorption. The chiral bands formed at the edges of a hybrid system made of a  quantum-anomalous-Hall-insulator and a superconductor yield equally spaced peaks in the CD signal. We identified the particle-hole 
selection rules responsible for this behavior from the analysis in terms of chiral bands.
In a disc there is no CD signal due to the incompatibility of the selection rules with the angular momentum restriction; a square or rectangular geometry (or, more generally, a system with straight edges) is needed.
The presence of two chiral bands can be inferred from the small splitting of the CD peaks.  Finally, both positive and negative CD peaks can be seen, with a perfect alternation in a square and a favored sign 
in a long 2D ribbon geometry. 

On the whole, our results suggest the use of CD spectroscopy as a valuable 
probe of chiral Majorana states, complementing the evidences obtained
with electrical conductance measurements \cite{He294}. This may require the use of
an array of absorbing devices, in order to achieve a combined signal of sufficient intensity.
Alternatively, techniques such as those developed for single plasmonic nanoparticle sensing \cite{Olson} 
might be applied to an isolated chiral-Majorana device.

\begin{acknowledgements}
This work was funded by MINECO (Spain), grant FIS2014-52564.
\end{acknowledgements}

\bibliography{cirdic}

\end{document}